\documentclass[twocolumn,showpacs,aps,prb]{revtex4}
\usepackage{graphicx}
\usepackage{dcolumn}
\usepackage{bm}

\begin{document}

\title{Realization of a classical counterpart of a scalable design for adiabatic quantum computation.}


%\affiliation{Friedrich Schiller University, Department of Solid
%State Physics, D-07743 Jena, Germany}
\author{V. Zakosarenko}
\author{N. Bondarenko}
\author{S. H. W. van der Ploeg}
\author{A. Izmalkov}
\author{S. Linzen}
\author{J. Kunert}
\author{M. Grajcar}%
\altaffiliation[On leave from ]{Department of Experimental Physics,
Comenius University, SK-84248 Bratislava, Slovakia.}
\author{E.
Il'ichev} \email{ilichev@ipht-jena.de}
\author{H.-G. Meyer}
\affiliation{%
Institute for Physical High Technology, P.O. Box 100239, D-07702
Jena, Germany\\
}%

\date{\today}

\begin{abstract}
We implement a classical counterpart of a scalable design for
adiabatic quantum computation. The key element of this design is  a
coupler providing controllable coupling between two bistable
elements (in our case superconducting rings with a single Josephson
junction playing the role of a classical counterpart of
superconducting flux qubits). The coupler is also a superconducting
ring with a single Josephson junction that operates in the
nonhysteretic mode with a screening parameter of about 0.9. The
flux-coupling between two bistable rings can be controlled by
changing the magnetic flux through the coupler. Since the coupling
amplitude is proportional to the derivative of the coupler's
current-flux relation, the coupling can be tuned from ferromagnetic
to anti-ferromagnetic. In between the coupling can also be switched
off.
\end{abstract}

\pacs{85.25.Dq}% PACS, the Physics and Astronomy
\maketitle


The magnetic properties of a single-junction interferometer
(superconducting loop with one Josephson junction) depend on its
normalized critical current $\beta=2\pi LI_{C}/\Phi_{0}$ only. Here
$L$ is the loop inductance, $I_C$ the critical current of the
junction and $\Phi_{0}$ is the flux quantum. If $\beta > 1$ such an
interferometer exhibits a double degenerated energy state if an
external flux equal to half a flux quantum is applied (degeneracy
point) \cite{Barone}. Close to the degeneracy point, the single
junction ring is a bistable element with two magnetic moments
corresponding to the superconducting screening currents flowing
clockwise and counterclockwise. These two states can be described in
spin formalism by making use of Pauli matrices. In other words, a
system of magnetically coupled interferometers is a realization of a
two-dimensional Ising spin system. When the coupling between spins
is randomly distributed the system represents a spin glass. The
problem of finding the ground state of such a system is a
nonpolynomial one, i.e., the amount of calculation resources needed
grows exponentially with the number of elements. Mathematically this
task is equivalent to solving the so-called MAXCUT
problem.\cite{MAX}

%Thus we have classical system with ground state in which a
%non-polynomial problem is encoded. Unfortunately, classical system
%itself has problem to find such ground state, the state with
%minimal energy. This gives direct physical insight that classical
%systems (computers) cannot solve some problems efficiently.
\begin{figure}[htb]\centering
\includegraphics[width=0.8\linewidth]{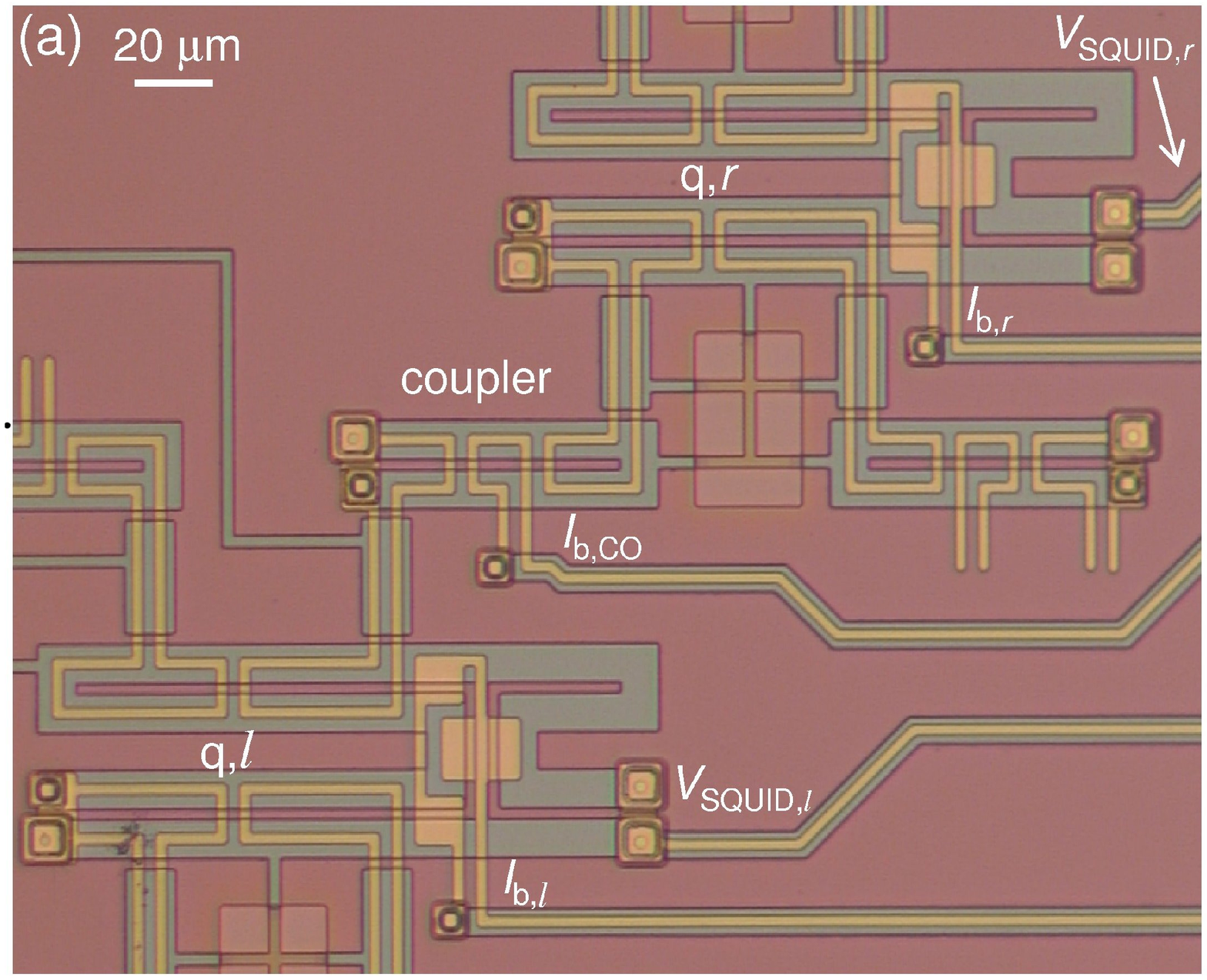}\\
\medskip
\includegraphics[angle=270,width=0.8\linewidth]{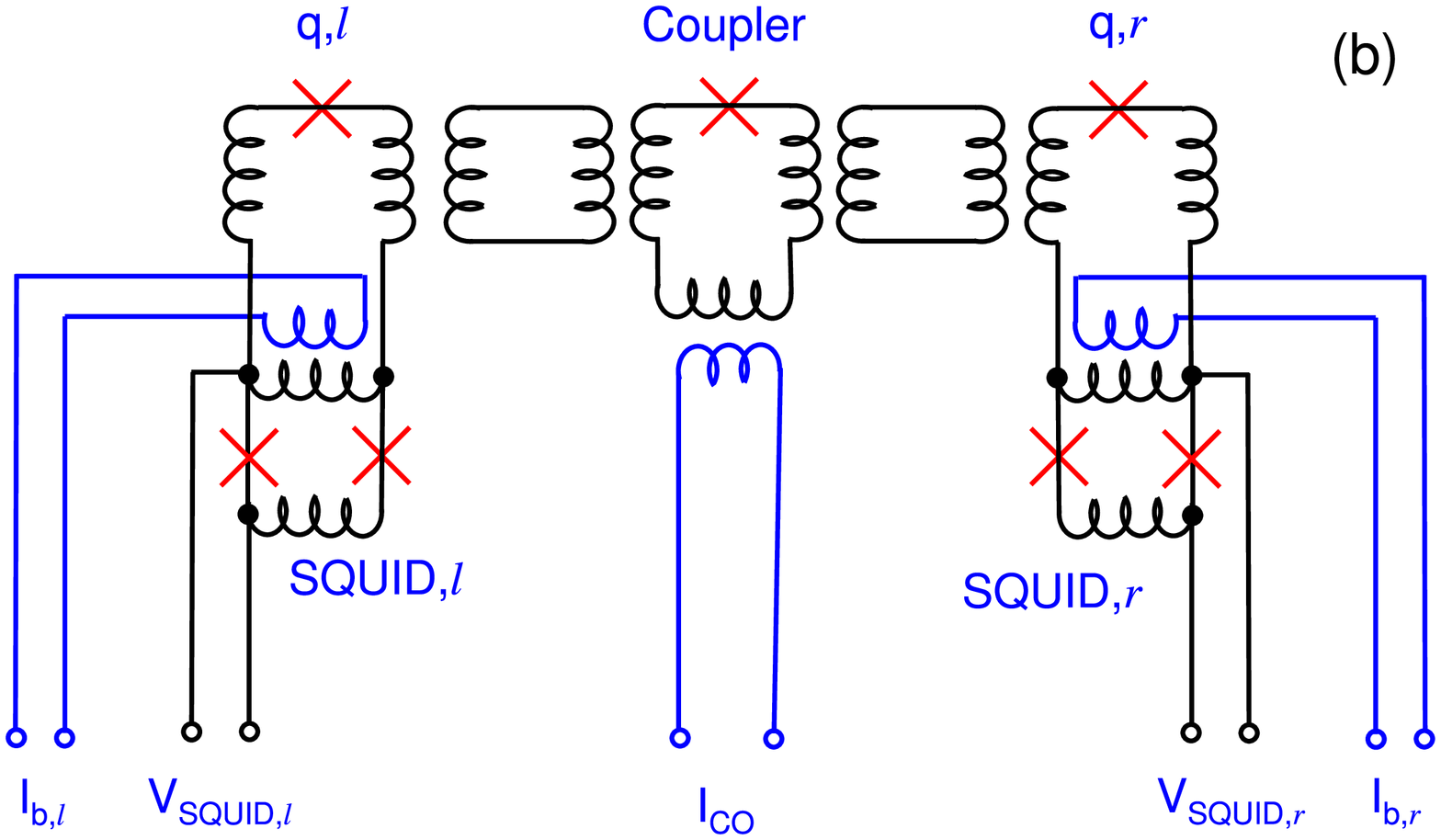}\\%viewport=0 0 550 480
\caption{(a) Photograph of the sample. (b) Circuit diagram. Two
bistable elements (q,l and q,r) are coupled to each other by the
coupler. The signals from bistable elements are read out by
\textit{left} and \textit{right} SQUIDs. The fluxes through the
interferometers could be controlled by the currents
($I_{\mathrm{b,l}},~I_{\mathrm{b,r}},~I_{\mathrm{CO}}$) through the
appropriate bias lines (b,l; b,r; and b,CO). All junctions are
nominally the same: $3.5\times3.5~\mu$m~$^2$.} \label{scheme}
\end{figure}

Recently, the implementation of an adiabatic quantum
algorithm\cite{Farhi} by making use of superconducting flux qubits has
been proposed. \cite{Kaminsky,Grajcar} Schematically, the
implementation is the same as for the 2D Ising model described above
but instead of classical bits quantum bits (qubits) are used. It has
been shown that such a quantum system could solve the problem
discussed above in polynomial time. In this letter we present the
realization of a scalable classical design implementing the 2D Ising
model with a coupling which can be tuned from antiferromagnetic,
through zero, to ferromagnetic.

Theoretically several schemes of tunable coupling were proposed
\cite{Clark, Filip} and also some
implemented.\cite{Castel,Ploeg2006} In the design presented here we
use the approach proposed by Maassen van den Brink \emph{et
al.}\cite{MaassenvdBrink2005} Here a coupler, realized as a
single-junction interferometer with $\beta_{CO}=2\pi
L_{CO}I_{C}/\Phi_{0}<1$, provides tunable coupling between two
bistable elements. The coupling amplitude $J$ can be varied by
changing the magnetic flux through the coupler. The value of $J$ is
proportional to the derivative of the coupler's current-flux
relation $I_s(\Phi_e)$, which presents itself as a periodic function
versus external magnetic flux $\Phi_e$. Therefore both
antiferromagnetic and ferromagnetic couplings can be realized
\textit{in situ}. The coupling between interferometers could also be
switched off when ${dI_s}/{d\Phi_e}=0$. We have experimentally
demonstrated both mentioned types of coupling and have been able to
switch the coupling off in between.

The scalable design is build up out of basic cells, each containing
one bistable element directly coupled to a readout DC-SQUID
(superconducting quantum interference device) and two equal couplers
coupled by flux transformers to the bistable element. In the basic
cell there are also flux bias lines for each coupler and one for
both the DC-SQUID and the bistable element. For the demonstration of
the basic working principle we used a structure containing two basic
cells of this scalable classical design. The samples were fabricated
using our standard technology with Nb-AlO$_x$-Nb
junctions.\cite{Stolz} The circuit diagram and a photograph of this
investigated structure are shown in Fig.\ref{scheme}.

The bistable elements (denoted by $q,l$ and $q,r$, see
Fig.\ref{scheme}) are designed as interferometers with
$\beta_{q}\sim 2.2$. The coupler is also designed as an
interferometer with $\beta_{CO}\sim 0.9$. Two bistable elements, are
coupled via a coupler and two flux transformers. The latter are used
in order to increase the coupling between $q,l$ and $q,r$ and
separate them in space. This resulted in reduction of unwanted
crosstalk between the flux bias lines. Each bistable element shares
a part of its inductance with the DC-SQUID used for its readout. The
value of this inductance is optimized in such a way that the biasing
of the DC-SQUID does not change the state of the bistable element. A
flux bias line coupled to this shared inductance can be used for the
adjustment of the external flux through the DC-SQUID and
interferometer as as well as for feedback in a flux-locked-loop
regime. All Josephson junctions are nominally the same ($3.5 \times
3.5~\mu$m$^2$). The desired screening factors for each element are
achieved only by the loop's geometry. The effective inductances of
the bistable elements, couplers, and DC-SQUIDs are designed taking
into account the screening due to the coupling with closed
superconducting loops. Their design values correspond to:
$L_{\mathrm{q}}=60$~pH, $L_{\mathrm{CO}}=25$~pH,
$L_{\mathrm{SQUID}}=50$~pH.

The mutual inductances between bistable elements, DC-SQUIDs,
couplers and appropriate bias lines were obtained from the
measurements of switching current histograms as a function of the
current through these lines. The mutual inductances found from these
measurements are: $M_{b,l;SQUID,l} \approx M_{b,r;SQUID,r}=5.4$~pH,
$M_{b,l;q,l} \approx M_{b,r;q,r} = 16$~pH and
$M_{b,\mathrm{CO};\mathrm{Coupler}}=6.4$~pH.

\begin{figure}[htb]
\centering
\includegraphics[width=0.85\linewidth]{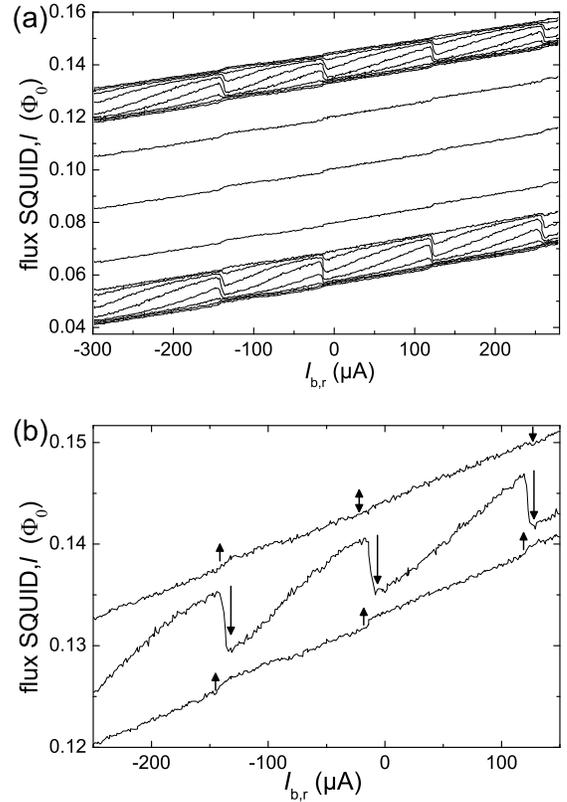} \centering
\caption{Demonstration of tunable coupling between two bistable
elements. (a) The output signal of the left DC-SQUID versus the
current $I_{b,r}$ applied to the right bias line. The curves
correspond to fixed bias current $I_{b,CO}=$ 524, 526, 528, 530,
532, 534, 536, 538 and 540; 600, 700 and 800; 850, 852, 854, 856,
858, 860, 862 and 864$~\mu$A (from top to bottom) producing dominant
magnetic flux in the coupler. The readout DC-SQUID operates in
flux-locked-loop mode. (b) Curves from the upper panel for coupler's
bias line current $I_{b,CO}=$526, 532 and 540~$\mu$A (from top to
bottom),  the first curve is close to the transition from
antiferromagnetic to ferromagnetic coupling. Since the magnetic flux
through the coupler weakly depends also on $I_{b,r}$, the gradual
crossover from antiferromagnetic to ferromagnetic regime is driven
by the right bias line current. The up, up-down, and down arrows
denote antiferromagnetic, ``zero'', and ferromagnetic couplings,
respectively.} \label{LSvsIbr}
\end{figure}
\begin{figure}[htb]
\centering \includegraphics[width=0.85\linewidth]{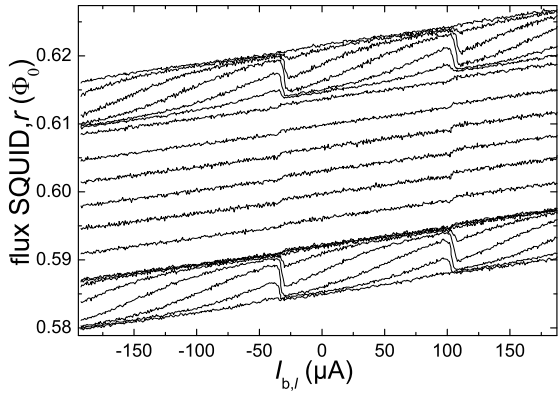}\\
\caption{Output signal of the right DC-SQUID versus the current
$I_{b,l}$ applied to the left bias line. The current trough the
coupler are from top to bottom $I_{b,CO}=$ 530, 528, 526, 524, 522,
520 and 518; 510, 460, 410, 360 and 310; 210, 208, 206, 204, 202,
200, 198 and 196$~\mu$A.}\label{RSvsIbl}
\end{figure}

For the characterization of the coupler we measured the current
induced in the right bistable element as a function of the current
circulating in the left one, \emph{i.e.} the response of the right
DC-SQUID to the left bias-line current (see Fig.\ref{scheme}) and
\textit{vice versa}. For the measurements the DC-SQUID was shunted
with an external resistor soldered onto the sample holder. The
DC-SQUID was operated in flux-locked-loop mode making use of
standard room temperature electronics \cite{Supracon}. The
electronics amplifies the DC-SQUID voltage and supplies the feedback
current through the appropriate bias line, keeping the total flux in
the DC-SQUID constant. For our design it means that the output
signal of the DC-SQUID electronics is proportional to the screening
current in the interferometer which is roughly proportional to the
flux coupled to it because $\beta_{q}>1$. The measurements were
carried out in liquid He at 4.2 K.

The coupling strength $J$ between the interferometers $q,l$ and
$q,r$ was tuned by applying a magnetic flux $\Phi_e$ through the
coupler using its bias line $\propto I_{b,\mathrm{CO}}$ (see
Fig.\ref{scheme}). The value of $J$ increases with the coupler's
screening factor $\beta_{\mathrm{CO}}$. On the other hand,
$\beta_{\mathrm{CO}}<1$ is required in order to obtain the change of
the coupling sign. Thus, for optimal results we need a sample with
$\beta_{\mathrm{CO}}\lesssim 1$. Taking into account the coupler
inductance $L_{\mathrm{CO}}=25$~pH, we fabricated samples with the
junction oxidation process for $j_c=100~A/cm^2$ giving a critical
current for the junction $I_C\approx 12~\mu$A and
$\beta_{\mathrm{CO}}\approx 0.9$ close to unity.

The results are shown in Fig.\ref{LSvsIbr} were the left DC-SQUID is
used as a measuring device and in Fig.\ref{RSvsIbl} for measurements
with the right DC-SQUID. The slope of the curves is due to the
direct parasitic coupling between the right (left) bias line and the
left (right) bistable element. The periodical modulation of these
curves is due to the jumps of screening current of the right (left)
interferometer. Note that the period of this modulation is the same
as the period measured with the DC-SQUIDs directly coupled to the
interferometer confirming that we are looking at the effect of this
interferometer.

In Figs.\ref{LSvsIbr} and \ref{RSvsIbl} the curves with upward and
downward kinks correspond to antiferromagnetic and ferromagnetic
couplings, respectively. The rounding is due to an averaging of the
thermally activated  jumps between two stable branches of the
interferometer. The kink height is proportional to the derivative of
the couplers' current phase relation: $dI/d\Phi_e=2\pi I_C \cos
\phi/[\Phi_0(1+\beta_\mathrm{CO}\cos \phi)]$ with
$\phi=2\pi\Phi_e/\Phi_0-\beta_\mathrm{CO} \sin \phi$ the phasedrop
over the coupler junction. So the maximal kink heights in the ferro-
and antiferromagnetic cases are proportional to
$\beta_\mathrm{CO}/(1-\beta_\mathrm{CO})$ and
$\beta_\mathrm{CO}/(1+\beta_\mathrm{CO})$, respectively. The
observed ratio of the maximal kink height for these two cases 12:1
corresponds to $\beta_\mathrm{CO}=0.85$ which is in reasonable
agreement with the expected value. Also, the fact that the
ferromagnetic coupling region is smaller than the anti-ferromagnetic
one qualitatively agrees with this picture. However, the
experimentally observed width of this region is only $0.04~\Phi_0$
while this simple model would predict $0.23~\Phi_0$ for
$\beta_\mathrm{CO}=0.85$.
%But qualitatively this simple model does not describe the width of this regions.
%The experimentally observed width of the ferromagnetic region is much smaller than the model predicts.
This could be due to a non-sinusoidal current phase relation of the
coupler junction. Further experiments would be necessary to clear up
this disagreement. In between the ferro- and the antiferromagnetic
regions the coupling vanishes as indicated by the updown arrow.

In conclusion, we demonstrate tunable coupling between two single
junction interferometers in classical mode. We show both
ferromagnetic and anti-ferromagnetic type of coupling, including
zero coupling between these two regimes. This system enables to
study relaxation and annealing in spin glass.

S.v.d.P., A.I., E.I. were supported by the EU through the RSFQubit
and EuroSQIP projects and M.G. by Grants VEGA 1/2011/05,
APVT-51-016604 and the Alexander von Humboldt Foundation. We thank
H.~E.~Hoenig for fruitful discussions. We also thank J. Hilton for
comments.

Partial financial support by the D-Wave Systems Inc. is gratefully
acknowledged.

%\newpage

\end{document}